\documentclass[runningheads]{llncs}
\usepackage{graphicx}
\usepackage{times}
\usepackage{amsmath}
\usepackage{amssymb}

\begin{document}
	
	\title{Generalized Deep Learning-based Proximal Gradient Descent for MR Reconstruction}
	
	 \author{Guanxiong Luo\inst{1} \and
	 Mengmeng Kuang\inst{2}\and
	 Peng Cao\inst{2}}
	 \institute{University Medical Center Göttingen\and The University of Hong Kong}

	\maketitle  
	\begin{abstract}
		The data consistency for the physical forward model is crucial in inverse problems, especially in MR imaging reconstruction. The standard way is to unroll an iterative algorithm into a neural network with a forward model embedded. The forward model always changes in clinical practice, so the learning component's entanglement with the forward model makes the reconstruction hard to generalize. The deep learning-based proximal gradient descent was proposed and use a network as regularization term that is independent of the forward model, which makes it more generalizable for different MR acquisition settings. This one-time pre-trained regularization is applied to different MR acquisition settings and was compared to conventional $\ell_1$ regularization showing \textasciitilde 3 dB improvement in the peak signal-to-noise ratio. We also demonstrated the flexibility of the proposed method in choosing different undersampling patterns.
		
	\end{abstract}

\section{Introduction}
Before employing deep learning in accelerated MRI reconstruction, conventional methods for parallel MR imaging are based on the numerical pseudo-inversion of ill-posed MRI encoding matrix, which could be prone to reconstruction error at poor conditioning \cite{pruessmann2001advances,lustig2007sparse,griswold2002generalized}. 
The encoding matrix comprises the k-space under-sampling scheme, coil sensitivities, Fourier transform. The traditional reconstruction involves some gradient descent methods for minimizing the cost function of the k-space fidelity and the regularization term \cite{lustig2007sparse,uecker2014espirit}.
There is a tradeoff between the image artifact level and the under-sampling rate as limited by the encoding capacity of coil sensitivities. Nevertheless, the parallel imaging technique robustly provides the acceleration at factor 2-4 \cite{pruessmann2001advances,griswold2002generalized}. 
The compressed sensing technique exploits the sparsity property of MR images in a specific transform domain, such as the wavelet domain, in combination with the incoherent under-sampling in k-space, which enables even larger acceleration factor.

With the fast growth of machine learning, the supervised learning have been applied to MRI reconstruction \cite{NIPS2016_6406,schlemper2017deep,aggarwal2018model}. Those methods MRI encoding matrices were fully included in the neural network models. These models were trained with predetermined encoding matrices and corresponding under-sampling artifacts. After training, imaging configurations, including the k-space under-sampling schemes and coil sensitivities, associated encoding matrices, must also be unchanged or changed only within predetermined sampling patterns, during the validation and application, which could be cumbersome or to some extent impractical for the potential clinical use. 

To tackle this design challenge, we unroll proximal gradient descent steps into a network and call it Proximator-Net. Inspired by \cite{chen2020compressive}, the proposed method was adapted from proximal gradient descent. This study's objective was to develop a flexible and practical deep learning-based MRI reconstruction method and implement and validate the proposed method in an experimental setting regarding changeable k-space under-sampling schemes.


\section{Method - Proximal network as a regularization term}

\begin{figure}[ht]
	\centering
	\includegraphics[width=0.6\textwidth]{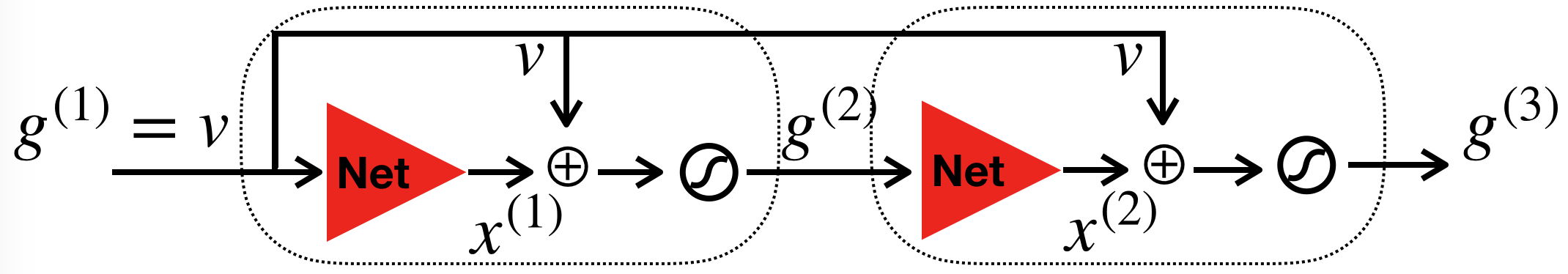}
	\vspace{-0.2cm}
	\caption{The unrolled proximate gradient descent as neural network, the $\mathbf{Net}_{\phi}$ are the red triangles whose detailed structure is detailed in supplementary material.}
	\label{net}
	\vspace{-0.5cm}
\end{figure}

The image reconstruction for MR k-space data acquired with different k-space trajectories can be formulated as an inverse problem for the corresponding forward model. We followed the MRI reconstruction problem formulation used in $l_1$-ESPIRiT \cite{uecker2014espirit}. Let a forward operator  $\mathcal{A}$ to map the MR image $\mathbf{x}$ to the sampled k-space data $\mathbf{y}$. The operator $\mathcal{A}$ consists of a Fourier transform operator $\mathcal{F}$, coil sensitivity $\mathcal{S}$, and a k-space sampling mask $\mathcal{P}$, i.e., $\mathcal{A}=\mathcal{PFS}$ \cite{uecker2014espirit}.The well-known solution to the inverse problem can be formulated as an optimization problem with a regularization term \cite{uecker2014espirit} as
\begin{equation}
\bar{\mathbf{x}} = \arg\min_x \|\mathcal{A}\mathbf{x}-\mathbf{y}\|_2^2+\lambda \phi(\mathbf{x}),
\label{eq:cs_model}
\end{equation}
where $\phi(\mathbf{x})$ is a regularization term that introduces the prior knowledge, and $\lambda$ the regularization parameter. The proximal operator for $\phi(\mathbf{x})$ in the proposed approach, $\mathbf{prox}_\phi\left(\mathbf{x}\right)$, is defined as:
\begin{align}
	&\hat{\mathbf{x}}={\mathbf{prox}}_{\phi,1/\lambda}\left(\mathbf{v}\right):=\arg\min_\mathbf{x}\|\mathbf{x}-\mathbf{v}\|_2^2+\lambda \phi(\mathbf{x}) 
	\label{eq:2a}\\
	&\hat{\mathbf{x}}\in\chi~ \sim P_{model}\approx P_{data} 
	\label{eq:2b}
\end{align}
where $\hat{\mathbf{x}}$ is the proximate value at $\mathbf{v}$, $\chi$ is the sub-space that contains MR images (including the ground truth), $P_{data}$ is the probability distribution of the observed MR images, and $P_{model}$ is learned the probability distribution.
Taking the sparse constraint in compressed sensing as an analogy, $\phi(\mathbf{x})$ is a $\ell_1$-norm of coefficients from the wavelet transform, the Equation \eqref{eq:2a} can be approximated by the shrinkage method (FISTA). Back to the definition of the proximal operator, we treat Equation \eqref{eq:2a} as another optimization problem solved iteratively by the proximal gradient descent method, which is expressed as follows
\begin{subequations}
	\begin{align}
	&\mathbf{g}^{(t+1)} = \mathbf{x}^{(t)} + \alpha^{(t)}(\mathbf{v}-\mathbf{x}^{(t)})\\
	&\mathbf{x}^{(t+1)} = {\mathbf{Net}}_{\phi}(\mathbf{g}^{(t+1)})
	\end{align}	
\end{subequations}
then, we unroll above iterative steps as a Proximator-Net shown in Figure \ref{net}.

At the iteration $t$, $\mathbf{x}^{(t+1)}$ is supposed be closer to $P_{data}$ than $\mathbf{x}^{(t)}$. In order to achieve this, images corrupted by Gaussian noises are used as input. The corresponding noise-free images are used as labels. The $\ell_2$-loss between outputs and labels is used. To solve Equation \eqref{eq:2b} and make the prediction close to the data distribution, the $\ell_2$ regularization of the gradient of output with respect to input is 
used, as proposed in \cite{alain2014regularized}. Therefore the final training loss is 
\begin{equation}
\mathbb{E}\left[\|r_\theta(\tilde{\mathbf{x}})-\mathbf{x}\|_2^2+\sigma^2\left|\left|\frac{\partial r_\theta(\mathbf{x})}{\partial\mathbf{x}}\right|\right|_2^2\right],
\end{equation}
where $\tilde{\mathbf{x}}$ is the image perturbed by the Gaussian noise with $\sigma$ and  $r_\theta(\mathbf{x})$ is the neural network parameterized by $\theta$.
Proximator-Net is recast as a neural network-based proximal operator for the proposed regularization term, and the proposed approach for reconstruction is
\begin{subequations}
	\begin{align}
	&\mathbf{m}^{(t)} = \mathbf{x}^{(t)} + \alpha^{(t)}\mathcal{A}^{H}(\mathbf{y}-\mathcal{A}\mathbf{x}^{(t)})\\
	&\mathbf{x}^{(t+1)} = (1-\lambda)\cdot\mathbf{x}^{(t)}+ \lambda\cdot{{r}}_{\theta}(\mathbf{m}^{(t)}) 	\label{eq:5b}
	\end{align}
	\label{eq:pro}
\end{subequations}
where $\lambda$ is the parameter for the learned regularization term. 

\section{Experiments and Results}
\textbf{Dataset and pre-processing:} The dataset we used in this work is from Ref. \cite{Luo_Magn.Reson.Med._2020}. Training images were reconstructed from 8 channels k-space data without undersampling. Then, these image data-sets after coil combination were scaled to a magnitude range of $[-1,1]$ and resized to an image size of 256$\times$256. In the end, 900 images were used for training, and 300 images were used for testing.  We set the level of Gaussian noise with $\sigma=0.03, \mu=0$. The training was performed with Tensorflow on 4 GTX 2080Ti GPUs and took about 2 hours for 500 epochs. Real and imaginary parts of all 2D images were separated into two channels when inputted into the neural network.
\vspace{-0.5cm}
\begin{table}[ht]
	\centering
	\caption{Comparison of PSNRs (in dB, mean $\pm$ standard deviation, N = 300) and SSIMs(\%) between $\ell_1$-ESPIRiT and the proposed method.}
	\label{tab:table_metrics}
	\vspace{-0.25cm}
	\includegraphics[width=0.9\textwidth]{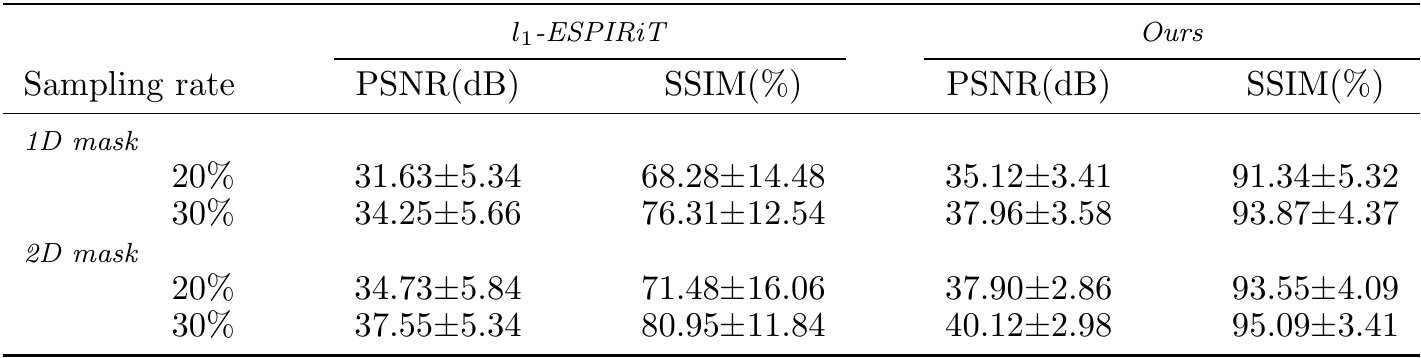}
	\vspace{-0.65cm}
\end{table}

\begin{figure}[ht]
	\centering
	\includegraphics[width=0.7\textwidth]{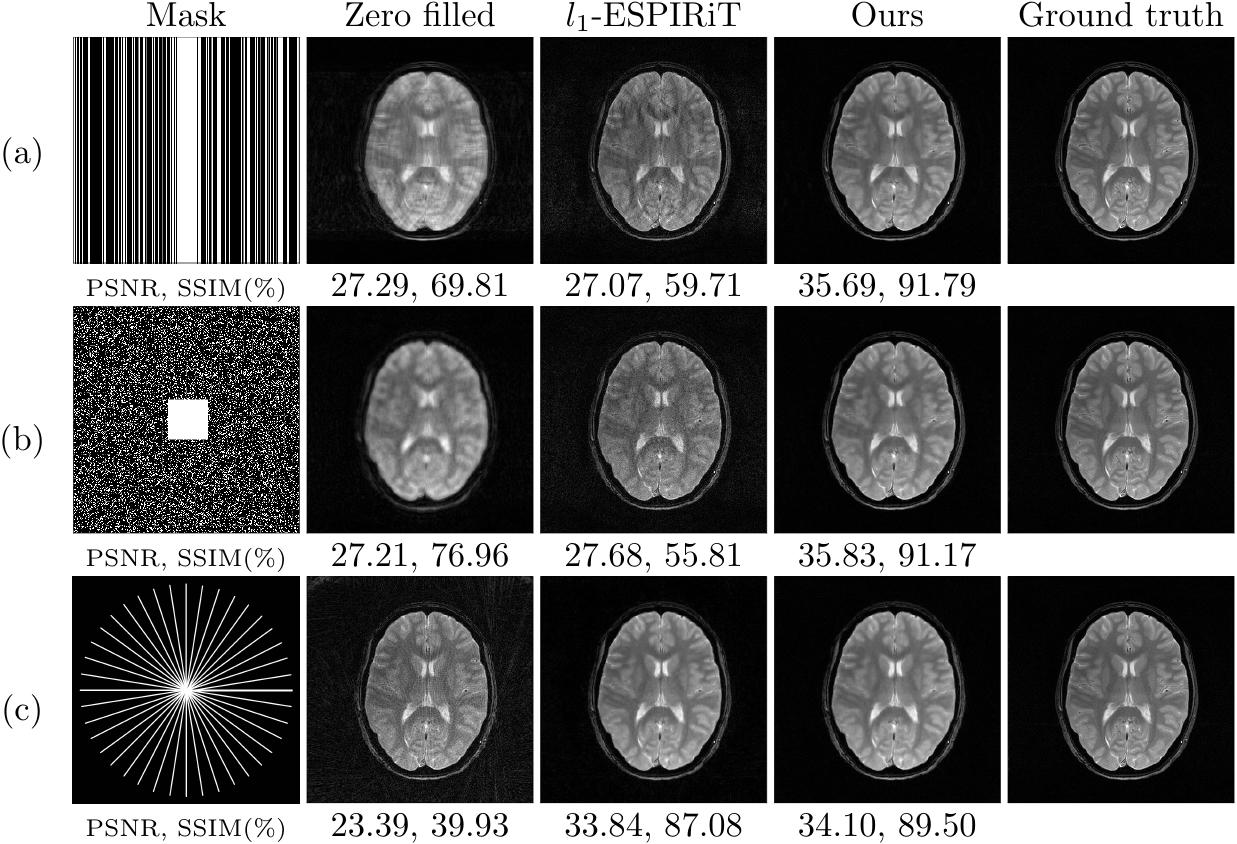}
	\vspace{-0.3cm}
	\caption{Row (a) shows the proposed method can remove the aliasing artifacts along phase encoding direction in the case of using 30\% k-space.  Row (b) shows the proposed method can eliminate the burring fog on the image in the case of using 20\% k-space. Row (c) shows the two methods' performances were close. With 40 radial spokes acquired in k-space.}
	\label{fig:ourresults}
	\vspace{-0.7cm}
\end{figure}

\noindent
\textbf{Reconstruction and comparisons:} The proposed method in Equation \ref{eq:pro} was implemented with Python. We performed $l_1$-ESPIRIT (regularization parameter=0.005) reconstruction with the BART toolbox \cite{uecker2014espirit,uecker2015berkeley}.
The coil sensitivity map was estimated from a 20$\times$20 calibration region in the central k-space using ESPIRiT. 
To compare the images reconstructed using different methods, we calculated the peak signal-to-noise ratio (PSNR) and structural similarity index measure (SSIM) with respect to ground truth. Also, we validated the proposed method in different acquisition settings such as acceleration along one dimension, two dimensions, or radial acquisition. The metrics of different reconstructions were shown in Table \ref{tab:table_metrics}.
Generally, the result demonstrated that the proposed method could restore the high-quality MRI with both high acceleration factor (i.e., 20\% samples) and high PSNR (i.e., $>$35 dB). Figure  \ref{fig:ourresults} shows the comparison between $l_1$-ESPIRIT and the proposed method.
\vspace{-0.5cm}

\begin{figure}[ht]
	\centering
	\vspace{-0.3cm}
	\includegraphics[width=0.68\textwidth]{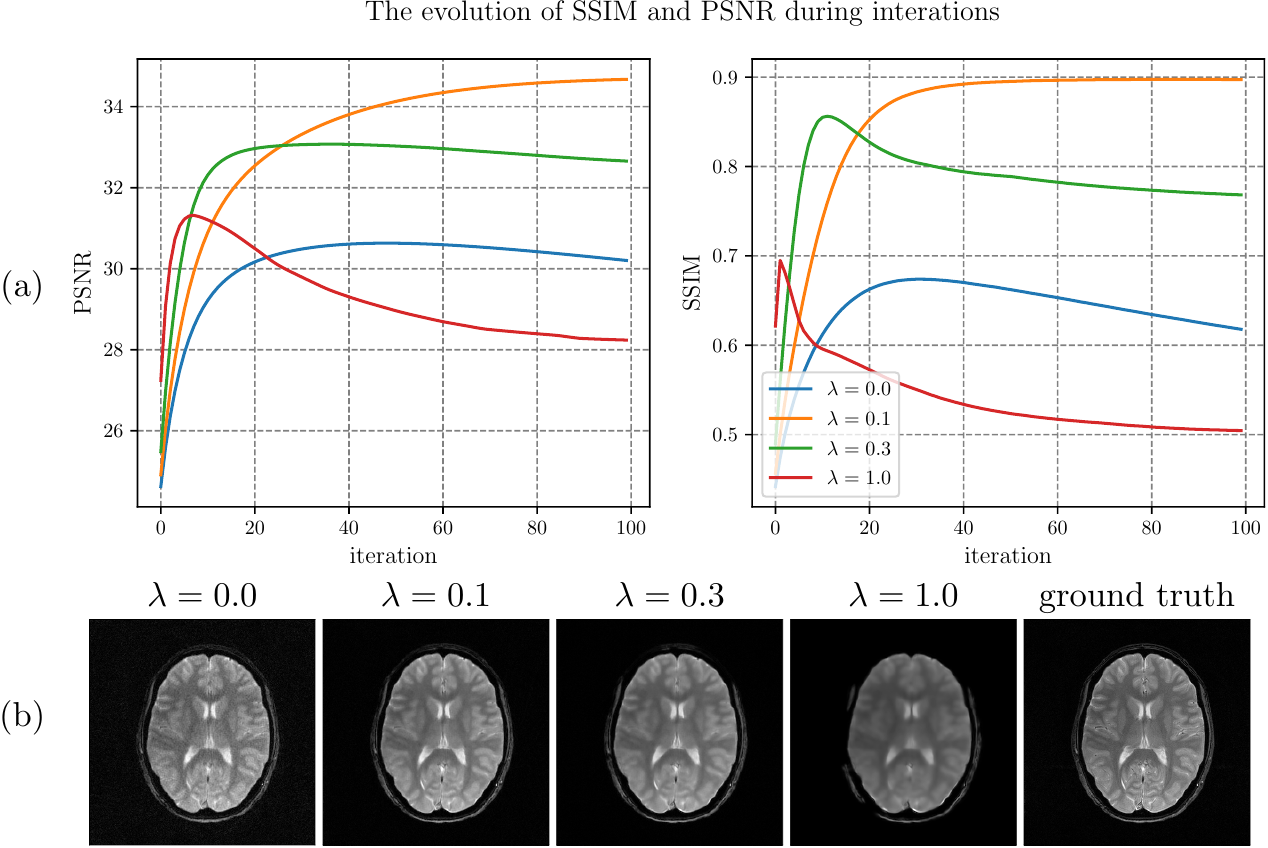}
	\vspace{-0.32cm}
	\caption{(a) The curve of SSIM and PSNR over iterative steps. (b) Images reconstruction from 40 radial k-space spokes with different $\lambda$.}
	\label{fig:curve}
	\vspace{-0.67cm}
\end{figure}

\noindent
\textbf{Smoothing effect of the learned regularization term:} 
The smoothing effect of the proposed regularization term was shown in Figure \ref{fig:curve}. The the parameter $\lambda$ was tunable in Equation \eqref{eq:5b}. When $\lambda$ was set to be zero, it was an iterative parallel imaging (i.e., SENSE) reconstruction \cite{pruessmann2001advances}. With the increase of $\lambda$, the artifacts and noises disappeared, and also the image appeared more smooth with some details eclipsed. We noticed that setting $\lambda$ around 0.1 provided better PSNR and SSIM.
We plotted SSIM and PSNR over iterations to monitor the quality of reconstruction, using varied tuning parameter $\lambda$.  

\section{Discussion and Conclusion}
In this study, a once pre-trained neural network - Proximator-Net - is used as a regularization in iterative algorithm for MRI reconstruction and can be applied to different reconstruction tasks, taking advantage of the separation of the learned information from the fidelity enforcement. That makes it different to previous methods \cite{NIPS2016_6406,schlemper2017deep,aggarwal2018model}. In this initial experiment, we focused on demonstrating the utility of the proposed method in classic compressed sensing and radial k-space acquisition, and we used the brain MRI data to evaluate the method. Like conventional iterative reconstruction algorithms, k-space fidelity in the proposed hybrid approach was enforced by the least-square term and implemented outside the neural network, allowing the high flexibility to change k-space under-sampling schemes and RF coil settings. For quantitative comparison, our methods achieved 3dB higher PSNR in the tested acquisition settings compared with $l_1$-ESPIRiT.

	\bibliographystyle{splncs04}
	\bibliography{references}

\end{document}